\documentclass[preprint2]{emulateapj}

\usepackage{amssymb}
\usepackage{multirow}

\usepackage{graphicx}

\def\cm{\,{\rm cm}}

\def\ergscm2 {erg\,s$^{-1}$cm$^{-2}$}

\def\cm2 {cm$^{-2}$}

\def\aap {A\&A}
\def\apj {ApJ}

\shorttitle{LSQ14bdq and SN 2006gz  : QNe in massive binaries}
\shortauthors{Ouyed et al.}

\begin{document}

\title{Quark-Novae in massive binaries :\\ a model for  double-humped, hydrogen-poor, superluminous Supernovae}
 
\author{Rachid Ouyed\thanks{Email:rouyed@ucalgary.ca}, Denis Leahy and Nico Koning}

\affil{Department of Physics \& Astronomy, University of Calgary, 2500 University Drive NW, Calgary, AB T2N 1N4, Canada}

\begin{abstract}
LSQ14bdq and SN 2006oz are super-luminous, hydrogen-poor, SNe with double-humped light curves.  
We show  that a Quark-Nova  (QN; explosive transition of the neutron star to a quark star) occurring in a massive binary, experiencing two Common Envelope (CE)  phases, can quantitatively  explain the light curves of LSQ14bdq and SN 2006oz. 
The  more massive component (A) explodes first as a normal SN, 
 yielding a Neutron Star which ejects the hydrogen envelope of the companion when the system
 enters its first CE phase.  During the second CE phase, the NS spirals into and 
 inflates the second He-rich CE. In the process it  gains mass and triggers a Quark-Nova,
outside of the CO  core, leaving behind a Quark Star.   
The  first hump in our model is the QN shock  re-energizing  the expanded He-rich CE.  
 The QN occurs when the He-rich envelope is near maximum size ($\sim 1000R_{\odot}$)  and  imparts
enough energy to unbind and eject the envelope. 
Subsequent merging of the Quark Star with the CO core of component B,  driven by gravitational radiation, 
   turns the Quark star  to a Black Hole.  The ensuing Black Hole  accretion provides sufficient power  for  the second brighter and long lasting hump.  Our model suggests a possible connection between SLSNe-I and type Ic-BL SNe which occur
    when the Quark Nova is triggered  inside the CO core.  We estimate the rate of QNe in massive
    binaries during the second CE phase  to be $\sim 5\times 10^{-5}$ of that of core-collapse SNe.
\end{abstract}

\keywords{circumstellar matter Ñ stars: evolution Ñ stars: winds, outflows Ñ supernovae: general Ñ supernovae: individual (LSQ14bdq,SN 2006oz)}


\section{Introduction}

Superluminous, hydrogen-poor, supernovae (SLSNe-I) reach peak luminosities at least
an order of magnitude higher than those of  Type-Ia SNe and standard core-collapse
SNe (e.g. \cite{galyam_2012}).  They show a variety of lights curves and  lack hydrogen in their spectra  (\cite{postorello_2010,quimby_2011}).  
They are associated with  low-metallicity environments/galaxies and
seem to appeal to different conditions than those of standard core-collapse and Type-Ia supernovae
(e.g. \cite{chen_2013}; \cite{nicholl_2015a};  \cite{leloudas_2015}; \cite{lunann_2015}).   
 Popular models include powering by a millisecond spinning-down pulsar with an $\sim 10^{14}$ G 
 magnetic field (\cite{woosley_2010, kasen_2010}) and  Black Hole (BH) accretion (\cite{dexter_2013}).  However, there is still
much debate about the progenitors and the mechanism/engine powering SLSNe-I.
  
Here we focus on two double-humped, hydrogen-poor, SLSNe namely 
LSQ14bdq (\cite{nicholl_2015b}) and  SN 2006oz (\cite{leloudas_2012}).
They  have  a precursor  (the first hump) which is much brighter and narrower than a typical SN
 and a second brighter and long lasting hump.     With two double-hump examples it is clear now 
 that the first hump is not a regular SN. Thus we revise our previous model for
 SN 2006oz (\cite{ouyed_2013}) in the context of the current work.
 
 Here we show that a Quark-Nova (QN) occurring in a massive binary undergoing two common
 envelope (CE) phases offers ingredients that can 
 account for the double-humped light curves and other general properties of SLSNe-I such
 as their hydrogen-poor nature. The paper is organized as follows: in \S 2 we give a brief overview of the QN model
 and the occurrence of QNe in binary systems. In \S 3 we show the results of applying our model 
  to LSQ14bdq and SN 2006oz. We provide a  discussion in \S 4 and conclude  in \S 5.

\section{Quark Nova Model}

The QN is the explosive  transition of a neutron star (NS) to a quark star (QS) (\cite{ouyed_2002, keranen_2005}; see  \cite{ouyed_2013a} for a review).   The released energy ejects on average 
 of $M_{\rm QN} \sim 10^{-3}$ M$_\odot$ of neutron-rich material (\cite{keranen_2005, ouyed_2009, niebergal_2010}). The relativistic QN ejecta  has   an average Lorentz factor $\Gamma_{\rm QN}\sim 10$
and a kinetic energy, $E_{\rm QN}$,   exceeding $10^{52}$ erg.  The QN can occur following  the explosion
of a massive single star or in a binary system.  The explosion is triggered when the parent NS reaches deconfinement densities in its core 
(e.g. \cite{staff_2006}) by becoming massive enough, via fall-back during the SN explosion, 
 accretion from a companion or from accretion while the NS is inside a CE as considered here.  
 The critical NS mass  above which the QN explosion is triggered  is defined hereafter as  $M_{\rm NS, c.}$
  which we take to be $2M_{\odot}$ to account for the most massive known NS  (\cite{demorest_2010}).

\subsection{Quark-Novae  in single-star systems}

A dual-shock QN (dsQN) happens when the QN occurs days to weeks after the  SN explosion
of the progenitor star. The time delay means 
that the QN ejecta catches up  and collides with the SN ejecta after it has expanded to large radii (\cite{leahy_2008, ouyed_2009}). 
Effectively,  the QN re-energizes the extended SN ejecta causing a re-brightening of the SN.  For  time delays 
  not exceeding a few days,   the size  of the SN ejecta is  small enough that only  a modest re-brightening results when the QN ejecta 
  collides with the preceding SN ejecta; this yields a moderately  energetic, high-velocity,  SN.  In  this case, however, the QN model predicts that the interaction of the QN neutrons with the SN ejecta
   leads to unique nuclear spallation products (\cite{ouyed_2011a})
   which may  have been observed   (e.g. \cite{laming_2014}). 
      For longer  time-delays,  extreme re-brightening occurs when the  two ejecta collide yielding  light curves very similar to 
   those of  SLSNe \citep{ouyed_2012, ouyed_2013b,kostka_2014}.  For time-delays exceeding many weeks, the SN ejecta is too large and diffuse to experience any substantial re-brightening.  The dsQN model has been  used to fit 
a  number of  superluminous  and double-humped supernovae (see {\it http://www.quarknova.ca/LCGallery.html} for a picture
gallery of the fits).

\subsection{Quark-Novae  in binaries}

 A QN could also occur in tight binaries where the NS can accrete/gain enough mass to  reach $M_{\rm NS, c.}$ and experience a QN  event.    The NS can  accrete either  from the  companion overflowing its Roche Lobe (\cite{ouyed_staff_2013})
 or while  inside a CE. 
QNe  in binaries  have proven successful
 in fitting  properties of unusual SNe.   For example, in a NS-WD system the detonation of a disrupted CO White Dwarf by the QN
 leads to an explosion  resembling SN 2014J (what
we referred to as  a QN-Ia;  \cite{ouyed_staff_2013, ouyed_2014, ouyed_2015}).  QNe in NS-(He)WD binaries have also been considered  (\cite{ouyed_2011b, ouyed_2011c}). 
In general,   a QN in binaries  provides:
 
 (i)   A means  (the QN ejecta) to shock, and reheat  the disrupted companion.
 The heating  is on the order of $\sim 10^9$ K for an ejecta mass
 of a few solar masses and QN energy of $\sim 10^{52}$ ergs which  in some cases
  may  trigger nuclear burning. 
     
(ii)   The QS is born with a magnetic field on the order of  $B_{\rm QS} \sim 10^{15}$ G owing to 
color ferromagnetism in quark matter during the transition (\cite{iwazaki_2005}).  
The spin-down (SpD) power from the QN compact remnant (the QS), thus provides an additional energy
source besides radio-active decay (if nuclear burning is induced by the QN shock).  This additional energy
source can be released on timescale  $\tau_{\rm QS}=4.74\ {\rm days}\ P_{\rm QS, 10}^2 B_{\rm QS, 15}^2$
  where the QS period is given in units of 10 milliseconds and its surface magnetic field in units of 
  $10^{15}$ G.  The spin-down power is important in QNe-Ia where in addition to runaway thermonuclear CO burning, the spin-down energy from the QS can   
   contribute to the light curve with interesting   implications
to Cosmology if some SNe-Ia are indeed QNe-Ia   (\cite{ouyed_2014, ouyed_2015}).

(iii)  The QS could convert to a BH and release additional energy via BH accretion
 (\cite{ouyed_2011b, ouyed_2011c}).  In a QN-Ia, the QS (a gravitational point mass) can slow down and trap some of the ejecta
 (e.g. \cite{ouyed_2015}).
 
(iv)  In general,  the QN explosion (if asymmetric) can provide a kick to the QS. For a  relativistic ($\Gamma_{\rm QN}\sim 10$)  QN ejecta
of mass $M_{\rm QN}\sim  10^{-3}M_{\odot}$, the QN kick is of the order of  a
100 km s$^{-1}$ for a 10\% asymmetry in the QN explosion.

 \begin{figure}
\centering
\includegraphics[scale=0.051]{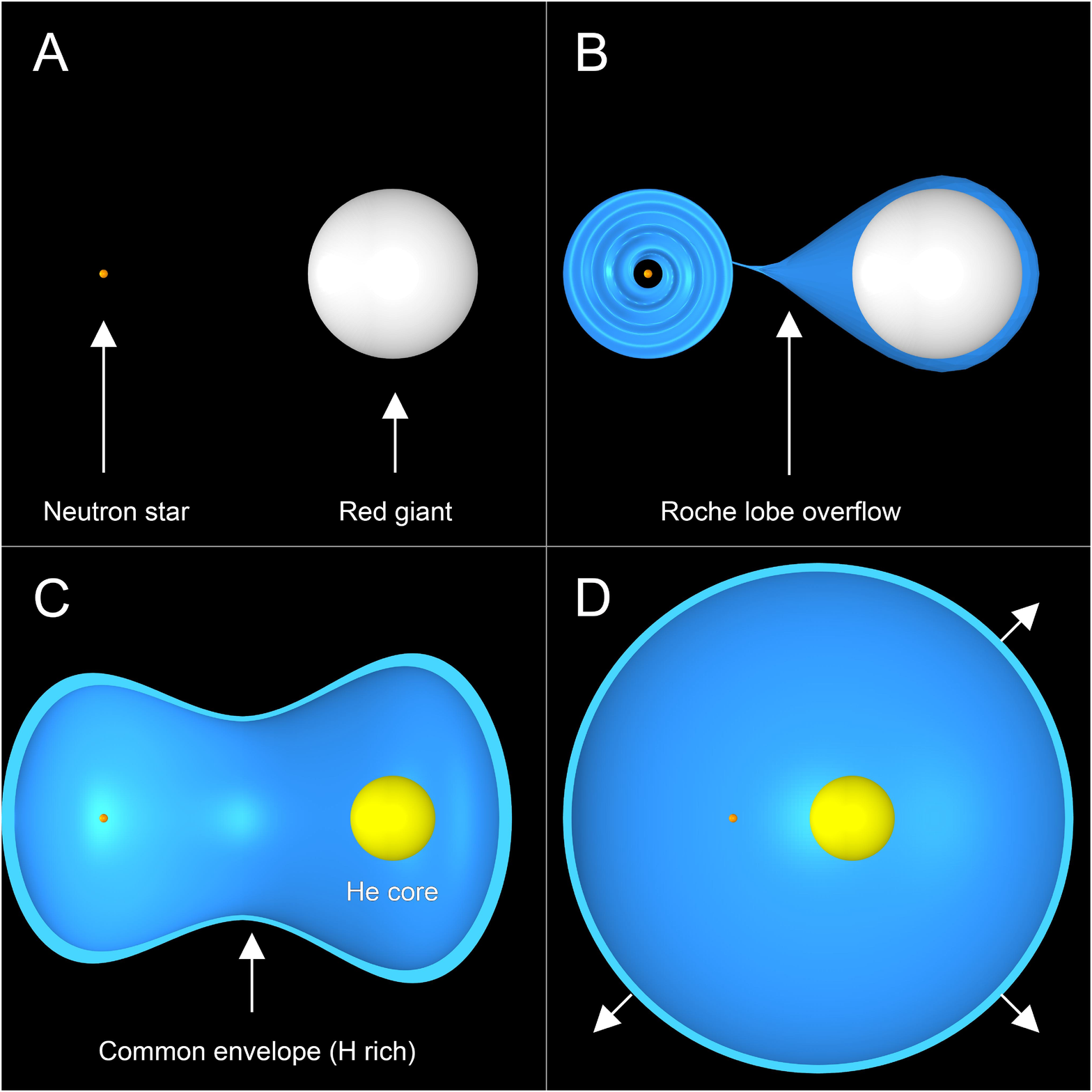}
\includegraphics[scale=0.051]{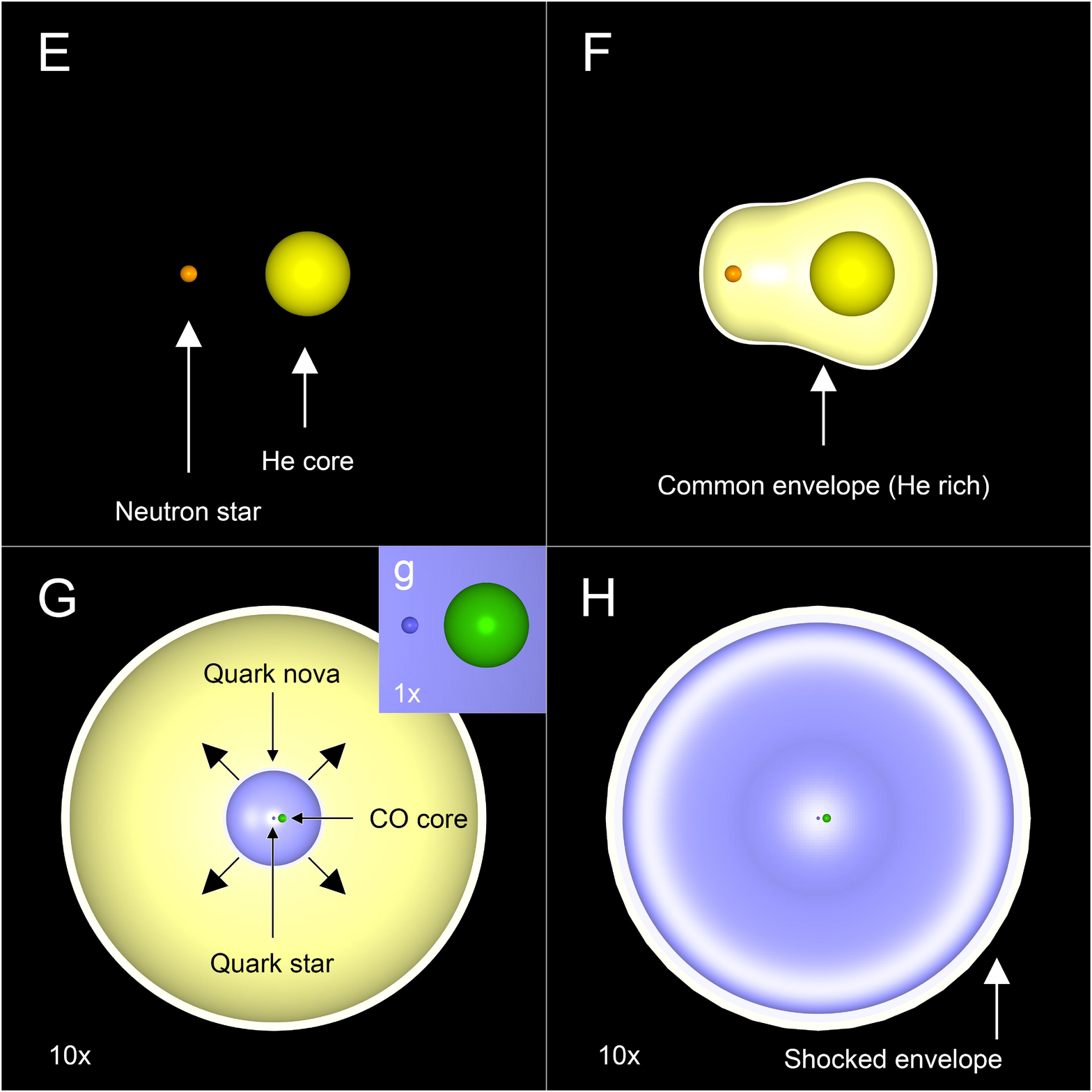}
\includegraphics[scale=0.051]{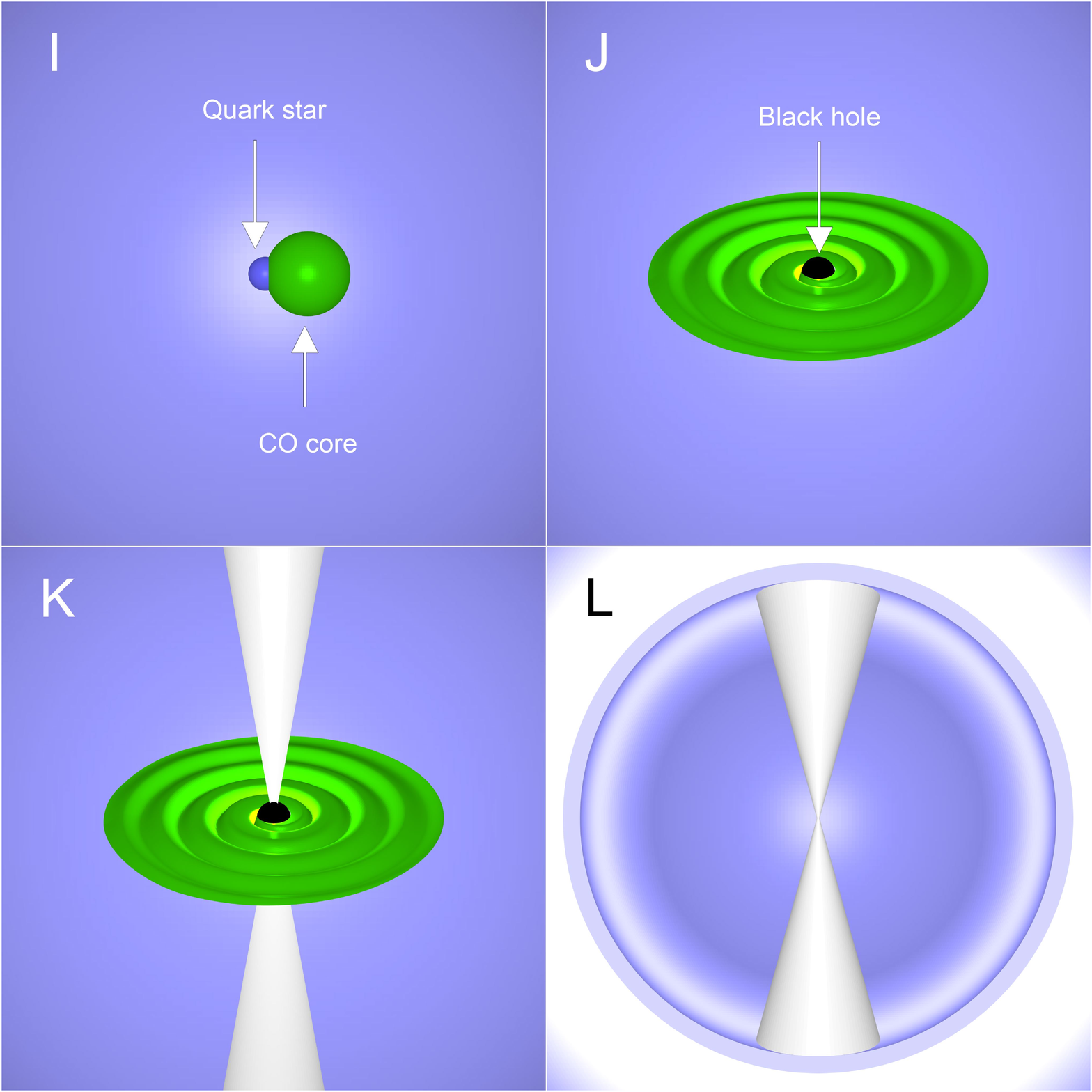}
\caption{Sequence of events in a massive binary leading to the QN event (leaving
behind a QS) and subsequent BH accretion
 of the core of the secondary companion following CE ejection (the BH forms following accretion onto the QS).  The
  ``1x" (``10x")  in panel g (G) stand for  1 (10) times  zoom.}
\label{fig:sequence}
\end{figure}

\section{The Quark-Nova in a massive binary : {\it Sequence of events}}
\label{sec:qnia}

Here we consider a QN   occurring  in a massive binary with both components  in the 
$\sim 20$-$25M_{\odot}$ range, a mass ratio close to unity and a binary
separation such that a first CE phase avoids merger (see \cite{taam_2010} and \cite{pod_2014} for a review).
We also assume a low-metallicity environment.   The following sequence of events 
leads to the QN explosion inside the second He-rich CE (see panels in Figure \ref{fig:sequence}):

(i)  The more massive component (A) experiences an SN explosion leaving behind a NS  with mass at birth
$M_{\rm NS, b.}\sim 1.4M_{\odot}$. 
  We only consider cases where the binary survives which is possible  considering a mass ratio
 close to unity (panel A).    A typical initial state is shown in Panel ``A"  which consist of 
 a NS and a Red Giant separated by $\sim 200 R_{\odot}$.
 
(ii) When component B evolves to the giant phase with a well developed Helium (He) core, the system
 enters the first CE envelope phase  leading to the ejection of the hydrogen envelope of B by the 
 in-spiralling NS (panels B, C, and D).

In the red giant phase, a component B of mass $\sim 20M_{\odot}$  for example would have
a radius large enough to  engulf the NS (panels ``C" and ``D") with a He core mass and radius of $\sim 6.0 M_{\odot}$
and $\sim 0.5R_{\odot}$, respectively.  In comparison, a  component B of mass $25M_{\odot}$ 
would have  a core mass and radius of $\sim 8.0 M_{\odot}$
and $\sim 0.8R_{\odot}$, respectively (e.g. Table 7.1  of \cite{salaris_2005}).
 In this phase, the density gradient  is steep enough (e.g. \cite{taam_1978}) to reduce the accretion much below
 the Hoyle\&Lyttleton rate (\cite{hoyle_1939}). 
  Efficient CE ejection is expected with at most   $\sim 0.1M_{\odot}$ of
  mass accreted by the NS (e.g. \cite{macleod_2015}) reaching a mass of $\sim 1.5M_{\odot}$.
  Thus to reach $M_{\rm NS, c.}\sim 2M_{\odot}$, the NS would need to accrete the remaining mass during the second CE
  phase as described below.
  
 (iii)  As illustrated in panel E,    following the first CE phase, one is left with a He-core-NS binary in a close orbit (e.g. \cite{hall_2014}; see also Figure 4 in \cite{macleod_2015})
  with a period of a few hours. Specifically, the binary separation following ejection of the first CE we estimate to be  of the order of
  $a_{\rm i} \sim 3R_{\odot}$ for an $M_{\rm NS, b.}\sim 1.4M_{\odot}$  NS  and a He core  of $\sim 6M_{\odot}$-$8M_{\odot}$
  representative of the 20-25$M_{\odot}$ mass range of the binary components.

  The  secondary B continues to evolve and expands to cause a second CE phase (panel F). 
 For a high enough mass ratio, a runaway mass transfer is expected once the He-rich core overflows its
Roche Lobe  leading to the  second CE phase  (e.g. \cite{ivanova_2003, dewi_2003})\footnote{See \cite{tauris_2015}  for other outcomes.}  accompanying  further accretion onto the in-spiralling NS.

The secondary  would expand to a radius of $> 3R_{\odot}$ (thus
engulfing the NS).  At onset of the second He  CE phase, the envelope size is of the
order of  $3R_{\odot}$ compared to about $200R_{\odot}$  for the onset of the first CE. Thus the envelope density is much higher 
 (a factor $(200/3)^3\sim 10^5$) in the second CE phase leading to much higher accretion rates. 
  ($\sim 0.1M_{\odot}$ yr$^{-1}$; \cite{brown_1995, chevalier_1996}). Although the accretion luminosity
 is  neutrino-dominated in this   regime (e.g. \cite{chevalier_1996}), thermal energy is deposited in the CE by the photons
  at the Eddington rate of $L_{\rm Edd.}\sim 4\times 10^{38}$ erg s$^{-1}$ for a $\sim 1.5M_{\odot}$
  NS. Convection transfers the deposited energy rapidly to the CE surface layers where it is radiated
  away. This is the self-regulating mechanism as described in \cite{meyer_1979})
   in which  the frictional energy
  that is released during   spiralling-in  is transported to the surface by
  convection and is radiated away without causing ejection of the envelope.
  In our model,  the CE ejection is caused by the QN explosion which occurs
  when the CE is near maximum size. 
  
The maximum envelope size is defined by  $4\pi R_{\rm CE, max.}^2 \sigma T_{\rm CE, eff.}^4
  = L_{\rm Edd.}$ with the envelope's effective temperature  $T_{\rm CE, eff.}\sim 3000$ K typical
   of fully convective stars on a Hayashi track (\cite{hayashi_1961}). This yields $R_{\rm CE, max.}\sim 1000 R_{\odot}$ at which
  point the envelope stops expanding (i.e. a steady energy balance is reached). For a thermal
  expansion speed of $c_{\rm s, CE}=\sqrt{k_{\rm B} T_{\rm CE, eff.}/\mu_{\rm CE} m_{\rm H}}\sim 5$ km s$^{-1}$, the envelope settles at its maximum radius  
   within a few years ($\sim  R_{\rm CE, max.}/c_{\rm s, CE}\sim 4.5$ years); 
   $\mu_{\rm CE}=4/3$ is the  He envelope mean atomic weight, $k_{\rm B}$ the Boltzmann constant 
    and $m_{\rm H}$ the Hydrogen mass.

 (iv)    While the CE radius is evolving towards it maximum radius, the NS continues to in-spiral 
  and to accrete  enough mass  ($\sim 0.4M_{\odot}$
  in $\sim 4$ years) to reach the critical mass $M_{\rm NS, c.}\sim 2M_{\odot}$ and undergo an explosive
  transition to a QS (panel ``G");  the envelope radius at the time of the QN is $R_{\rm CE, QN}\sim  c_{\rm s, CE} \times (4\ {\rm years})
   \sim 900R_{\odot}$.    Interestingly the time it takes the NS to experience the QN event
  is  close to the time it would take the CE to reach its maximum size; I.e. $R_{\rm CE, QN}\sim  R_{\rm CE, max.}$
  thus offering a  picture where the QN energy is released when the CE is near maximum size.
   This provides conditions for optimum harness of the $\sim 10^{52}$ erg of QN kinetic energy which
   yields the bright first hump in our model. The energy released by the QN explosion is enough to unbind the CE
   and eject it. Thus   in our scenario (with a He-rich second CE), the QN helps 
   induce CE ejection.  The QN shock propagating at  velocity $v_{\rm QN, sh.}$ energizes the  He CE (panel H) and yields the
 first bright and short-lived hump;  the corresponding initial envelope temperature is estimated
 from shock physics (Lang 1999),  $k_{\rm B} T_{\rm CE, 0}\sim (3/16) \mu_{\rm CE} m_{\rm H} v_{\rm QN, sh.}^2$.
  There will be no nuclear/He burning induced by the QN shock since  the conditions in the envelope (temperature and density)
are  below critical values.

(v)  After CE ejection, one is left with a $\sim 2M_{\odot}$ QS and a CO core
  of mass $M_{\rm CO} \sim 2M_{\odot}$ and radius $ < 0.1R_{\rm sun}$. The orbital
  separation when the QN occurs can be derived by estimating by how much the NS
 orbital radius  has decreased when it experiences the QN event.
  For a Hoyle\&Lyttleton rate accretion rate, a relationship between the NS orbit and mass 
  can be derived which is $a_{\rm QN}\sim a_{\rm i} (M_{\rm NS, i}/M_{\rm NS, c.})^\sigma$
  with $5 \le \sigma \le7$ as given in \cite{chevalier_1993} (and references therein). For
  $\sigma\sim 6$, $M_{\rm NS, i}\sim 1.5M_{\odot}$ and $M_{\rm NS, c.}\sim 2M_{\odot}$ 
  we get $a_{\rm QN}\sim 0.18 a_{\rm i} \sim 0.54 R_{\odot} a_{\rm i, 3}$ where
  $a_{\rm i, 3}$ is the initial orbit separation just before the second spiral-in phase starts (i.e. 
  immediately after the first, H-rich, CE has been ejected) in units of $3R_{\odot}$.

  The orbit  then decays through emission of gravitational radiation.
General relativity predicts that decay of orbital period is given by (\cite{landau_1962})
\begin{equation}
\frac{dP}{dt} = - \frac{192\pi G^{5/3}}{5c^5} \left( \frac{P}{2\pi} \right)^{-5/3} M_{\rm NS} M_{\rm CO} M_{\rm T}^{-1/3} \ ,
\end{equation}
where $M_{\rm T}= M_{\rm NS}+M_{\rm CO}$ is the total mass and we assume
zero orbital eccentricity. After integrating 
equation above and making use of Kepler's third law,  this gives a decay timescale
\begin{equation}
\tau_{\rm GW}\sim 9.4\ {\rm days}\  \frac{a_{\rm i, 3}^4}{M_{\rm CO, 2}M_{\rm NS, 2}M_{\rm T, 4}}   \ ,
\end{equation}
where  $M_{\rm CO, 2}$, $M_{\rm NS, 2}$ and $M_{\rm T, 4}$ 
are the CO core mass, NS  mass and total mass  in units of $2M_{\odot}$, $2M_{\odot}$ and $4M_{\odot}$, respectively.
The time delay between the QN event and merging is very sensitive to the QS-CO-core separation 
 at the start of the second spiral-in phase ($a_{\rm i}$). To get two humps in the
 lightcurve, $a_{\rm i}$ can only vary by a factor of 2.

 (vi) The orbital period of the system when the QS reaches  the core is of the order
of a few minutes for a CO core of $< 0.1R_{\odot}$ in radius. Thus  once  the QS reaches the core, merging occurs  very quickly
  and on timescales not exceeding a few hours (i.e. in many orbital periods).
 The QS merges with the CO core of B  and accretes enough mass (a
fraction of a solar mass) to become a black hole (BH).  The CO core is spun-up
by the in-spiralling NS/QS (panel I). The bulk of the CO core forms a disk around the BH  and accretes onto it (panels J, K and L)
producing a luminosity $L_{\rm BH}$
to power the long lasting main hump of the light curve. 
The BH accretion phase  is delayed from the QN event by the time required for the QS to reach the CO core ($\tau_{\rm GW}$), turn into a BH,
merge with the core and trigger accretion. We
 define this time delay as $t_{\rm BH, delay}\simeq \tau_{\rm GW}$ at which point the CE has extended to a radius
 $R_{\rm CE, BH}=R_{\rm CE, QN}+ v_{\rm CE} t_{\rm BH, delay}$ with
  $v_{\rm CE}= (3/4)v_{\rm QN, sh.}$  the CE expansion velocity induced by the QN shock; recall
  that $R_{\rm CE, QN}$ is the CE radius at the onset of the QN.

\subsection{Case study : LSQ14bdq and SN 2006oz}

\begin{figure*}
\centering
\includegraphics[scale=0.6]{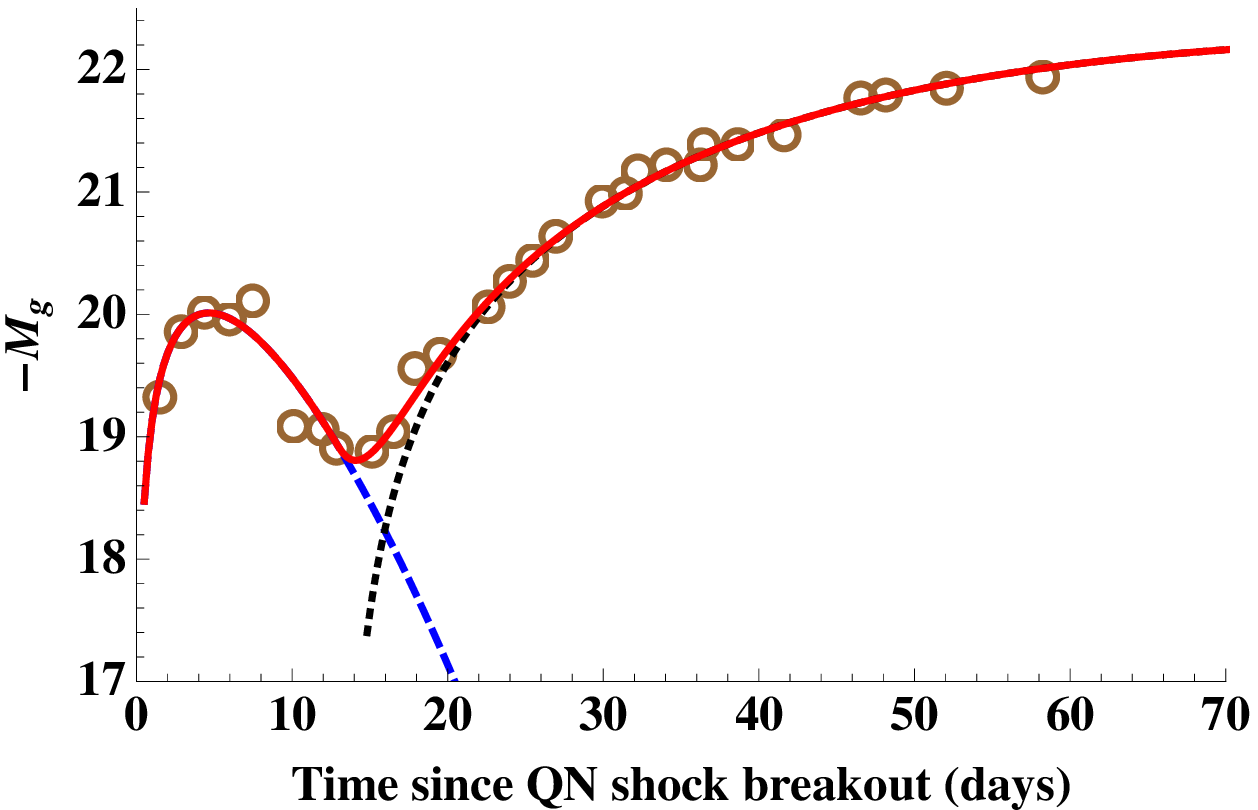}
\includegraphics[scale=0.6]{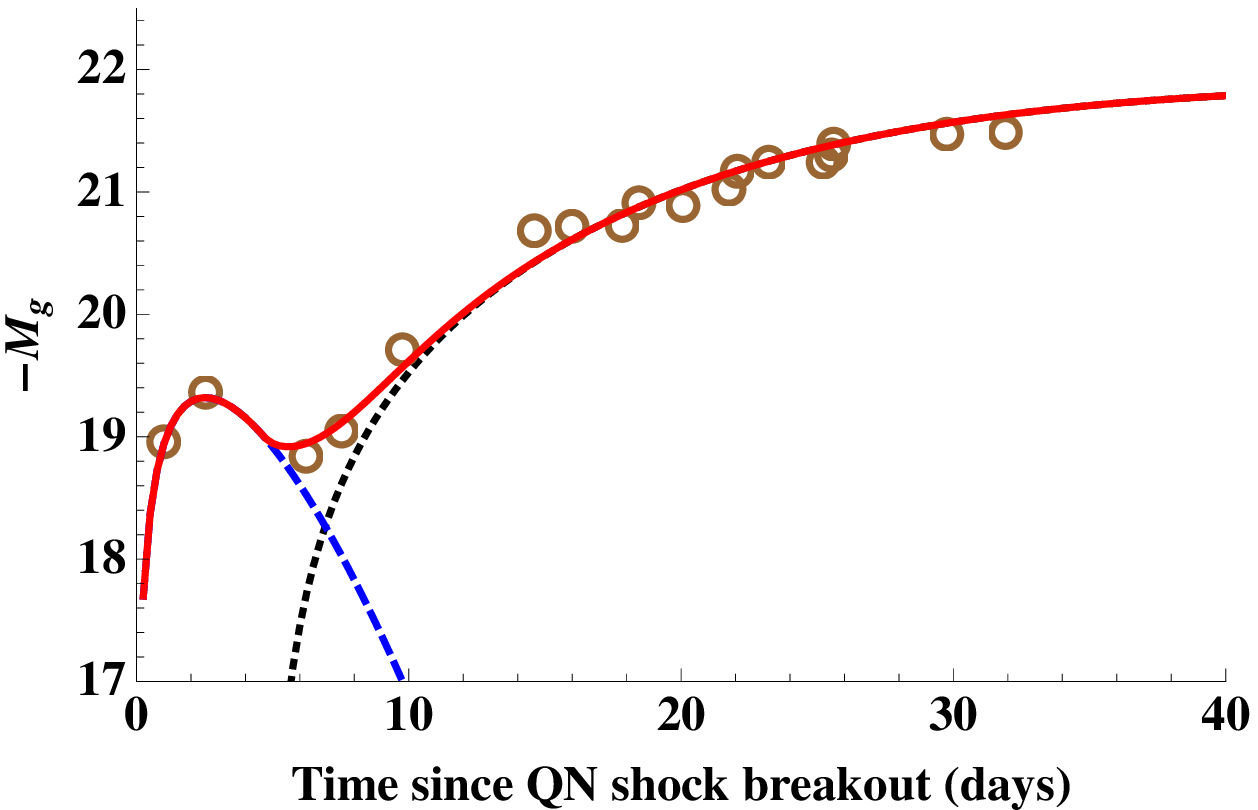}
  \caption{Left Panel: The QN model fit (solid line) to the g-band light curve of LSQ14bdq.  The observations (the circles) are from
 Nicholl et al. (2015b). Right Panel: The QN model fit (solid line) to the light curve of SN 2006oz.  The observations (the  circles) are
  from Leloudas et al. (2012).  In all panels, the blue dashed curve is the QN while the black dotted curve is the BH accretion.}
\label{fig:fits}
\end{figure*}

Table 1 shows the model's best  fit parameters for LSQ14bdq.
The left panel in Figure \ref{fig:fits} shows the resulting fit in the g-band absolute magnitude for LSQ14bdq together with 
 the observations  from \cite{nicholl_2015b}.  The blue dashed curve shows the QN proper (the first
 hump in our model) for   $M_{\rm CE}= 8M_{\odot}$ and $R_{\rm CE, QN}= 1050R_{\odot}$.  The black dotted 
 curve is from BH accretion with  an initial accretion
 luminosity $L_0=3\times 10^{44}$ erg s$^{-1}$.   We use the prescription in
 \cite{dexter_2013}   and adopt a  constant energy injection case
(see their Appendix)\footnote{The constant energy injection adopted  for the BH-accretion model is valid only for the
early stages of BH-accretion powered lightcurve (i.e. $< 50$ days; see Figure 2). A more realistic calculation
should include time-dependent energy injection. We do not fit the late tail which would require fully integrated
numerical light curves.}.   Table 1 also shows the model's best  fit parameters for SN 2006oz with the resulting model plotted in 
 the right panel in Figure \ref{fig:fits} together with  observations by \cite{leloudas_2012}.
 In Table 1, $E_{\rm QN}\sim E_{\rm CE, th.}+E_{\rm CE, K}$ where $E_{\rm CE, th.}$ for 
 simplicity was calculated ignoring the contribution of radiation which needs 
 a more detailed assessment.

The  duration of the  accretion
 power in our model is given by the CO core mass $M_{\rm CO} = M_{\rm B}- (M_{\rm H}+M_{\rm CE})$ divided by the
 accretion rate; $M_{\rm H}$ is the hydrogen lost by component B (of initial mass $M_{\rm B}$) during the first CE phase.    An $L_{\rm BH}\sim 10^{44}$ erg s$^{-1}$  corresponds to an accretion rate of
 $\sim 1M_{\odot}$ yr$^{-1}$ for an accretion efficiency of $\sim 10^{-3}$;  similar accretion rates onto the BH during CE  evolution have been found  by other studies (e.g. \cite{armitage_2000}; see also \cite{dexter_2013}). This gives
  the duration of the accretion powered phase of a  few hundred days in agreement with the duration of the
  observed long lasting second humps. The power from BH accretion was delayed by $t_{\rm BH, delay} = 12.5$ days for LSQ14bdq
and 4.5 days in the case of SN 2006oz.

 \begin{table}
\centering
\caption{Best fit parameters  for the LSQ14bdq and SN 2006oz LC  in our model.}
\scalebox{0.8}{
\begin{tabular}{cccccc} \hline
  & \multicolumn{2}{c}{He-rich (i.e. second) CE} &  \multicolumn{1}{c}{QN} & \multicolumn{2}{c}{BH Accretion} \\
  \hline
SLSN-I  &  $M_{\rm CE}$ ($M_{\odot}$) & $R_{\rm CE, QN} (R_{\odot})$    &  $v_{\rm QN, sh.}$ (km/s) &  $t_{\rm BH, delay}$ (days) & $L_{\rm 0}$ (erg/s)   \\\hline
 LSQ14bdq &  8  & 1050   & 30,000  & 12.5  &  $3\times 10^{44}$    \\\hline
  SN 2006oz &    4  & 500   & 40,000  & 4.5  &  $2\times 10^{44}$  \\\hline
\end{tabular}}\\
The corresponding  kinetic energies of the CE/ejecta, $E_{\rm CE, K}$   in units of  $10^{52}$ ergs  are: 
 $\sim 4$ for LSQ14bdq and $\sim  2.7$ for SN 2006oz. The thermal energy is $E_{\rm CE, th.}\simeq E_{\rm CE, K}$.
\label{table:params}
\end{table}

\subsection{Spin-down power}
\label{sec:spd}

\begin{figure}
\centering
\includegraphics[scale=0.6]{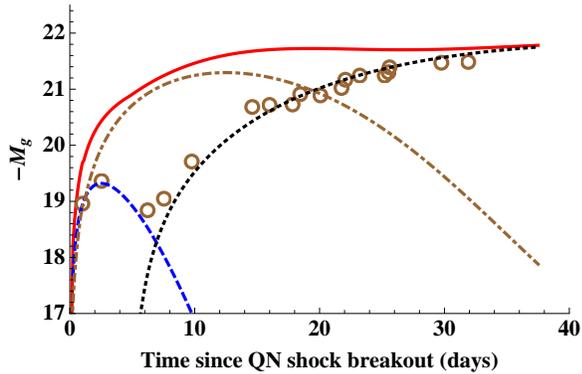}
  \caption{Effect of QS spin-down power (dot-dashed curve) on the g-band light curve of  SN 2006oz.  The
  combined light curve (solid line) is a single broad hump.}
\label{fig:SpD}
\end{figure}

  Following the QN-induced ejection of the second CE, accretion onto  the QS  is drastically reduced. The QS is free
  to spin-down (SpD) and to release up to $E_{\rm SpD}= 2\times 10^{52} {\rm ergs}\ P_{\rm QS, ms}^{-2}$ ergs in rotational energy;  the QS inherits its millisecond period from the  NS which was spun-up
  to millisecond equilibrium period (e.g. \cite{tauris_2012}).
   The QS magnetic field of  $\sim 10^{15}$ G (\cite{iwazaki_2005}) imply SpD timescale of $\sim 1.1\ {\rm hours}$.
   Figure \ref{fig:SpD} shows the resulting lightcurve for SN 2006oz with SpD included using the 
   prescription of \cite{woosley_2010,kasen_2010}.   As can be seen
   SpD  overwhelms the QN and when combined with BH-accretion the model yields
   a single  broad  hump.    The QN shock deposits energy throughout the CE on short timescales and is released by the
receding and cooling photosphere yielding a bright and narrow lightcurve. This is in contrast to the SpD (and also BH-accretion) where the energy
is deposited centrally and is thus affected by the envelope's diffusion timescale  which
 leads to a broad lightcurve.   

The good fits to the two observed double-humped SLSNe (see Fig. 2)
     suggest that either the SpD is negligible (e.g. for $M_{\rm NS, b.}\sim 1.8M_{\odot}$ the QN occurs
     before the NS is fully recycled) or that SpD power is highly beamed.

\section{Discussion}
\label{sec:discussion}

We have shown that  a QN in a massive binary experiencing 
two CE phases (in a low-metallicity environment) can  account for the 
double-humped light curves of the SLSNe-I LSQ14bdf  and SN 2006oz. 
We note the following:

 (i)  The QN is unlikely to occur in the first CE envelope since the NS can accrete at most
   $\sim 0.1M_{\odot}$ (e.g. \cite{macleod_2015}). However,
  if the NS is born massive (e.g. $M_{\rm NS, b.}\sim 1.9M_{\odot}$)  then a QN can occur in the first CE as well.
  In this case,  a ``purely" Hydrogen-rich SLSN may be the outcome and possibly 
  a double-humped lightcurve as described above if the QS also merges with the 
  He core.

(ii)     The QN hump has some universal
features since the QN energy $\sim 10^{52}$ erg and the envelope's
size when the QN occurs $R_{\rm CE, QN}\sim R_{\rm CE, max.}\sim 1000R_{\odot}$ (within a factor of $\sim$ 2)
 do not vary much.  However, if the 
delay between the QN and the onset of BH accretion ($t_{\rm BH, delay}$) is short, 
the QN hump will overlap with the BH-accretion powered hump in which case
the SN should appear as a single  long bright hump.

 (iii)  When the two humps are  distinguishable, the first hump should show spectral features of an He-rich ejecta (the CE envelope) while the second
 longer lasting hump should also show spectral signatures of the CO-core material (the BH jet) 
 when as the He envelope becomes optically thin.

 (iv)  In our model, the time delay between the QN (which powers the first hump) and BH formation
 and subsequent accretion (which powers the second hump)  is  the time it takes the QS to reach, and merge with, the 
  CO core.  However if the binary separation is such that $a_{\rm i}>> 3R_{\odot}$ following
  first CE ejection (i.e. $a_{\rm QN}>> 0.5R_{\odot}$ when the QN occurs) 
   then the merger timescale can be many years  making the QN event and the
   BH-accretion as  chronologically separate events. 
  Only one   hump (the QN) would be seen, yielding a type Ib  SLSN.
   The outcome is a  QS-CO-core binary which can evolve to
 a QS-WD or QS-NS binary depending on the conditions of the CO core.
 
 (v)    The BH accretion luminosity  in our model of $\sim 10^{44}$ erg s$^{-1}$ (i.e.
     BH accretion rates of  $\sim$ 0.1-1$M_{\odot}$ yr$^{-1}$ at $\sim 10^{-3}$ efficiency) is low compared to the 0.01-0.1$M{\odot}$ s$^{-1}$
     accretion rates of typical long duration   GRBs (e.g. \cite{piran_2004}). I.e. we do not expect a GRB during  the BH
     in  our model (see \ref{sec:bh-grb}).

(vi) Here we estimate the rate  of H-poor SLSNe from QNe which occur during
the second CE phase of massive binaries, including both single-
and double-humped events (i.e. all values of
$t_{\rm BH, delay}$).  For a Salpeter mass function (\cite{salpeter_1955}) we estimate about $\sim$ 10\% contribution to core-collapse SNe (CCSNe)
from stars in the $\sim 20$-$25M_{\odot}$ range.  With about $\sim 10$\% of these having a companion in the
20-25$M_{\odot}$ mass range this gives $f_{\rm binary}\sim 1$\% of binaries as considered here.
We take an estimate of those binaries that survive the 
 first SN explosion as $f_{\rm survival} \sim  20$\% (e.g. \cite{kalogera_1996}).  In a low-metallicity
 environment (with weak stellar winds), we expect  most 
 of those who survived the first SN explosion  to enter the first CE phase. The fraction
 of these binaries in a low-metallicity environment we take to be $f_{\rm Z} \sim$ 10\%
 (i.e. they are more likely to occur in galaxies such as the LMC and SMC than in the Milky Way).
 The $f_{\rm CE, 1}\sim$ 50\% of them which  survive the first CE phase (\cite{macleod_2015}) will undergo
 a second CE phase at a high rate of $f_{\rm CE, 2}\sim$ 100\%.
 Finally, we assume that in $f_{\rm QN}\sim 50$\% of cases the QN occurs before the merger wtih the CO core.
  The overall rate is then $\sim f_{\rm binary}\times f_{\rm survival} \times f_{\rm Z} \times f_{\rm CE, 1} \times  f_{\rm CE, 2}\times f_{\rm QN}  \sim   5\times 10^{-5}$.
 This is, within uncertainties, not very different from the observed rate of 
 H-poor, type Ic,  SLSNe  which are also very rare with a rate  of $\sim$ 3-8$\times 10^{-5}$
that of the core-collapse population (e.g. \cite{mccrum_2015}  and references therein).

 \subsection{A QN inside the CO core: {\it  A possible Type Ic-BL SNe and GRB connection}}
 \label{sec:bh-grb}
 
Many, but not all,  long duration GRB (LGRBs) are associated with Type Ic-BL SNe  (e.g. \cite{galama_1998, fynbo_2006}). Type Ic-BL SNe are more common than LGRBs;
 ``BL" stands for Broad-Line characterizing  their $\sim 30,000$-$40,000$ km s$^{-1}$ velocities. 
 In our model for SLSNe-I, the QN occurs during the second CE phase before the NS enters the CO core. 
 It is also possible for the QN to occur after the NS enters the CO core, for instance if the accreted mass
 in the He-rich CE is not enough to drive the NS above $M_{\rm NS, c.}$. 
 
 A QN  inside the CO core leads to the following events: 
 
 (i) It provides in excess of  $\sim 10^{52}$ ergs in energy which drives
   high-velocity CO-rich ejecta; $v_{\rm CO}\simeq 30, 000\ {\rm km\ s}^{-1} (E_{\rm QN, 52}/M_{\rm CO, 2})^{1/2}$.
    The compactness of the CO core means that most of the QN energy goes into the kinetic
    energy of the ejecta.    The CO ejecta is expanding inside the $\sim 1000R_{\odot}$ He envelope
    which is optically thick (optical depth of $\sim 10^5$). After the He envelope has expanded ($\sim$ 10-20 days) the high velocity CO ejecta
    becomes visible and should have characteristics similar to type Ic-BL SNe.

    (ii)      Because the QN  most likely occurs off-center, 
    the central densest parts of the CO core remains gravitationally bound.
     The  remnant core would  form an accretion disk around the QS.  
      \cite{ouyed_2005}  showed that accretion onto a QS can reproduce the activity
    (duration and intermittency) of GRBs\footnote{\cite{vogt_2004} found photon emissivities in excess of $10^{50}$ erg cm$^{-3}$ s$^{-1}$ in color
     superconducting quark matter.   The electromagnetic quark plasma frequency
     means that photon emissivity is suppressed for photons with energies below $\sim 23$ MeV. For thermalized
     photons of energy $\sim 3 T_{\rm QS}$ (with $T_{\rm QS}$ being the QS temperature) this means that photon emissivity is shut-off when the QS cools
     below $\sim 7.7$ MeV. \cite{ouyed_2005}   found  that intermittent accretion-ejection (heating-cooling) episodes  occur  naturally during accretion onto 
     the QS with  an activity (variability and duration) in $\gamma$-ray emission similar to observed light curves of GRBs.}
    from an accretion disk with a mass of a fews tenths of a solar mass.  
     The highly collimated outflow from the QS accretion (see \cite{ouyed_2005}) has to break through the
     He envelope which takes of $\sim 1000R_{\odot}/c\sim 2000$ s.  If accretion lasts long enough and 
     the jet is viewed along its axis, a QS-GRB would be seen.

    (iii)  The QS will likely transition to a BH leading to a BH-GRB  in the case of 
     high accretion rates  ($\sim$ 0.01-0.1$M_{\odot}$ s$^{-1}$) onto the BH.
     The BH-GRB would immediately follow the QS-GRB. As noted above,
      the combined duration of the two GRBs should exceed $\sim 2000$ s to break through the He envelope.
       Additionally, the QS-GRB$+$BH-GRB  would be observed  only if the  viewing conditions are ideal which would make this an 
extremely rare event.  An important caveat is that  to get GRBs, accretion rates from the CO core  must be much higher
 than those used to fit the second hump in the lightcurves of LSQ14bdq and SN 2006oz. 
 This suggests that GRBs require special  physical conditions in this case.

\section{Conclusion}

   QNe should be more common in binaries with NSs (in particular if born massive)   where the NS can gain sufficient  mass to reach the
 critical  mass $M_{\rm NS, c.}$,  to trigger quark deconfinement in their cores and go QN (\cite{staff_2006}).
It is no surprise therefore that QNe should be more common in massive binaries experiencing two CE phases
where  the NS can gain the necessary mass from two mass reservoirs  and go over  the $M_{\rm NS, c.}$ limit. 
 In our model, the QN occurs while the CE has expanded close to its maximum radius of $\sim 1000R_{\odot}$
yielding a near ``universal" QN (the first) hump in the resulting light-curve. The second hump attributed to
BH-accretion occurs  when the QS turns into a BH as it merges with the CO core  days following
the ejection of the He-rich CE.
It is also natural to expect QNe, as described here, to favor low-metallicity environments where the mass reservoirs
(the two CEs) would have retained more mass over time.   In high-metallicity environment, the stellar
 wind of component B can remove  its envelope before it causes a second CE.
   The ability of the QN model in binaries to fit SLSNe in general (see  {\it http://www.quarknova.ca/LCGallery.html})
and in particular  double-humped hydrogen-poor SLSNe-I,  as shown here, suggests that QNe
   may be part of binary evolution.     At a more fundamental level,
    one of the key finding of this work is that in some cases QNe in binaries can assist with CE ejection. This can be tested
    by including the QN in simulations of common envelope evolution.

\begin{acknowledgements}   
This work is funded by the Natural Sciences and Engineering Research Council of Canada. 
\end{acknowledgements}



\end{document}